\documentclass[preprint2]{aastex}

\usepackage{graphicx}

\DeclareGraphicsExtensions{.pdf,.png,.jpg}

\def\kms{km~s$^{-1}$}
\def\vlsr{$v_{\rm LSR}$}
\def\simgt{\lower.5ex\hbox{$\; \buildrel > \over \sim \;$}}
\def\msol{{$M_\odot$}}

\shorttitle{CO Observations of the Supernova Remnant 3C 434.1}

\shortauthors{Jeong et al.}

\begin{document}

\title{CO {\it \textbf{J}}$=$1$-$0 and {\it \textbf{J}}$=$2$-$1 Line Observations of \\
the Molecular Cloud - Blocked Supernova Remnant 3C434.1}

\author{Il-Gyo Jeong\altaffilmark{1}, Bon-Chul Koo\altaffilmark{1}, Wan-Kee Cho\altaffilmark{1},
Carsten Kramer\altaffilmark{2}, J\"{u}rgen Stutzki\altaffilmark{3}, Do-Young Byun\altaffilmark{4,5}}

\altaffiltext{1}{Astronomy Program, Department of Physics and Astronomy,
Seoul National University, Seoul 151-742, Republic of Korea;
igjeong@astro.snu.ac.kr, koo@astro.snu.ac.kr}
\altaffiltext{2}{Instituto Radioastronomia Milimetrica (IRAM), Av. Divina Pastora 7,
E-18012 Granada, Spain}
\altaffiltext{3}{I. Physikalisches Institut, Universit\"{a}t zu K\"{o}ln,
Z\"{u}lpicher Stra\ss e 77, 50937 K\"{o}ln, Germany}
\altaffiltext{4}{Korea Astronomy \& Space Science Institute,
Daejeon 305-348, Republic of Korea}
\altaffiltext{5}{Yonsei University Observatory, Yonsei University, Seongsan-ro 262,
Seodaemun, Seoul 120-749, Republic of Korea}

\begin{abstract}

We present the results of CO emission line observations
toward the semicircular Galactic supernova remnant (SNR) 3C434.1 (G94.0$+$1.0).
We mapped an area covering the whole SNR in the
$\rm{^{12}CO}~{\it J}=1$--0 emission line
using the Seoul Radio Astronomy Observatory (SRAO) 6-m telescope, and
found a large molecular cloud
superposed on the faint western part of the SNR.
The cloud was elongated along the north-south direction and
showed a very good spatial correlation with
the radio features of the SNR.
We carried out $^{12}$CO ${\it J}$ = 2--1 line observations of this
cloud using the K\"{o}lner Observatorium f\"{u}r
Sub-Millimeter Astronomie (KOSMA) 3-m telescope and found a region in which
the $^{12}$CO ${\it J}$ = 2--1 to ${\it J}$ = 1--0 ratio was high ($\sim$ 1.6).
This higher excitation,
together with the morphological relation, strongly
suggested that the molecular cloud was interacting with the SNR.
The systemic velocity of the molecular cloud ($-13$~\kms)
gave a kinematic distance of 3.0 kpc to the SNR-molecular cloud system.
We derived the physical parameters of the SNR based on this new distance.
We examined the variation of the radio spectral index over the remnant and
found that it was flatter in the western part, wherein the SNR was interacting
with the molecular cloud.
We therefore propose that 3C434.1 is
the remnant of a supernova explosion that occurred
just outside the boundary of a relatively thin, sheetlike molecular cloud.
We present a hydrodynamic model showing that its asymmetric radio morphology
can result from its interaction with this blocking molecular cloud.

\end{abstract}


\section{Introduction}

Supernova remnants (SNRs) appear usually spherical when they are young.
Historical SNRs such as Cassiopeia A, Tycho, and Kepler are good examples.
As they evolve, however, their morphology can deviate considerably from
sphericity by their interaction with the ambient medium.
If the SNR encounters, for example, a dense molecular cloud,
the SNR shell becomes
distorted. In contrast, if the remnant
crosses a diffuse bubble, then
part of the SNR shell could blow out or could
even be completely missing.
Therefore, from the morphology of SNRs, we can infer
their environments, which are often useful for understanding the nature and
evolution of SNRs.

\begin{figure*}[!t]
\includegraphics[width=1\textwidth]{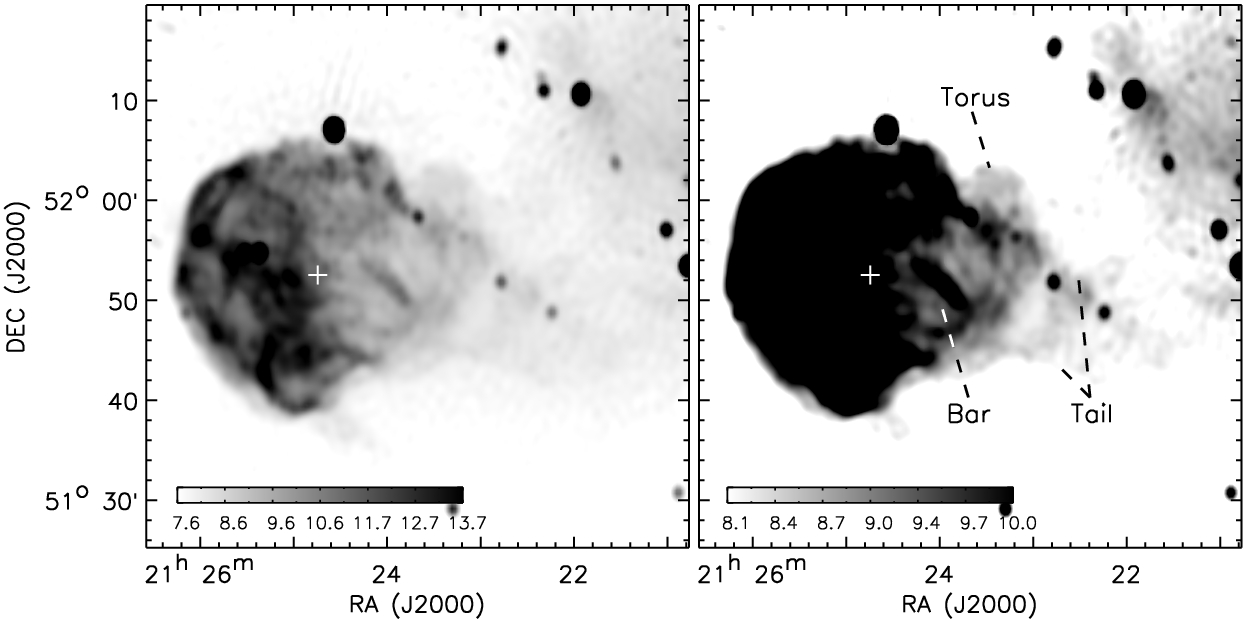}
\caption{Radio continuum image of 3C434.1 at 1420 MHz
from the Canadian Galactic Plane Survey data \citep{taylor03}.
The two frames show the same image with different gray scales to
emphasize the detailed radio continuum structure.
The central
white cross represents the geometrical center of the remnant.
The features discussed in the text are labeled.
\label{fig1}}
\end{figure*}

The SNR 3C434.1 (G94.0+1.0) is a shell-like supernova remnant with a
distinct morphology in radio
(\citealt{Willis73,Goss84,Landecker85,Foster05}; see Fig. {\ref{fig1}}).
It has a bright eastern hemisphere which has a well-defined, circular boundary.
The limb brightening indicates that we are seeing
the eastern half of a spherical shell with a radius of $13'$.
The complex filaments projected inside the remnant implies that the shell
is not uniform.
The circular shell-like structure continues
a little further to the west, but then
the morphology becomes quite complex.
Some noticeable features are as follows: (i) a
linear, convergent, thin ``tail" that originates
out from the western end of the
shell and extends much further ($\sim$ 20\arcmin) to the west,
(ii) a ``torus" that appears to be composed of several
ring structures surrounding the connecting portion of
the tail and the SNR shell,
and (iii) a $8'$-long ``bar" that is located near the center of
the torus and extends roughly parallel to the tail.
We discuss these features later in this paper.

\citet{Foster04} explored the atomic environment of the SNR using
Canadian Galactic Plane Survey (CGPS) HI 21 cm-line data. They
noted fragmented HI filaments along the SNR boundary and interpreted
them as parts of preexisting stellar wind bubble produced by the
massive progenitor of the SNR. They developed a dynamical model
for the SNR and its wind bubble based on this scenario.
The kinematic distance to the HI filaments (10 kpc), however, yielded
unreasonable parameters, and thus, they instead adopted the distance to
the Perseus spiral arm (5.2 kpc) as the distance to the SNR.
Later, \citet{Foster05} carried out follow-up multiwavelength studies
in radio to X-rays. He analyzed the archival ROSAT X-ray data to
show that the X-ray emission is bright and centrally-brightened in the eastern
part of the SNR and that the X-ray spectrum can be fitted with a
single-temperature thermal plasma model. The derived temperature and density
of the X-ray emitting plasma were $4.5\times 10{^6}$ K and 0.2 $\rm {cm}^{-3}$, respectively, with a revised distance of $4.5\pm 0.5$~kpc. He also detected
H$\alpha$ emission correlated with the radio structure
along the southeastern boundary of the SNR.

In this paper, we present the results of our CO study of 3C434.1 which gives a
picture of the SNR significantly different from the previous HI studies.
We found a large molecular
cloud in contact with the western boundary of the SNR.
The cloud was originally identified by
\citet{Huang86} in their low-resolution ($\sim 30'$)
CO survey of SNRs; however, no follow-up studies
have been made.
We present strong evidence for the interaction of the SNR
with the molecular cloud, and we argue that
the western part of the SNR is faint and complicated because
the remnant is interacting with the cloud there.
We derive the physical parameters of 3C434.1 based on
the new kinematic distance (3 kpc) and discuss
the variation of radio spectral index and the origin of
the distinct radio features.

\section{Observations}

$^{12}$CO ${\it J}$ = 1$-$0 (115.2712 GHz) observations were carried out from
November 2003 to December 2003 using the Seoul Radio Astronomy Observatory (SRAO)
6-m telescope. The FWHM beam size at 100 GHz was 120\arcsec
and its main
beam efficiency was 70\% \citep{Koo03}. We used a 100 GHz SIS mixer receiver
with single-side band filter and a 1024-channel auto-correlator with a 50 MHz
bandwidth for front-end and back-end, respectively.

We mapped the 41\arcmin\ $\times$ 39\arcmin\ (R.A. $\times$ Dec.) area centered at
(21$^{\rm h}$24$^{\rm m}$50$^{\rm s}$, 51\degr53\arcmin00\arcsec) with 1\arcmin~grid
spacing by using the position switching mode.
The velocity coverage was from $-$95 to $+$35~\kms.
The off-position was located at (21$^{\rm h}$22$^{\rm m}$28$^{\rm s}$,
51\degr40\arcmin00\arcsec).
The system temperature ranged from 500 to 800 K depending on the elevation and
weather condition. The typical rms noise level was $\sim$ 0.3 K at 0.5 \kms~
resolution. Each spectrum was integrated over 30 s. The relative calibration accuracy
had been checked by observing bright standard sources near the target every
one or two hours. Pointing accuracy was better than 20\arcsec.

$^{12}$CO ${\it J}$ = 2$-$1 (230.5380 GHz) follow-up line observations were performed in December 2004
using the 3-m telescope at the K\"{o}lner Observatorium f\"{u}r
Sub-Millimeter Astronomie (KOSMA). We used a dual-channel SIS receiver \citep{Graf98}
and acousto-optical spectrometers \citep{Schieder89} for front-end and back-end,
respectively. The beam width and main beam efficiency were 130\arcsec~and 68\%, respectively.
Pointing accuracy was better than 10\arcsec. We have mapped an area of
20\arcmin\ $\times$ 37\arcmin\ using the position-switched on-the-fly (OTF) mode with a grid
spacing of 50\arcsec. The OFF position was (21$\mathrm{^{h}}$35$\mathrm{^{m}}$39$\mathrm{^{s}}$.1,
52\degr42\arcmin59\arcsec). The velocity resolutions was 0.2~\kms\ and
the typical rms noise level was $\sim$ 0.3 K.

The data were reduced by using CLASS\footnote{http://www.iram.fr/IRAMFR/GILDAS}
software,
which is part of the Gildas package for single-dish spectral line analysis developed by
Grenoble and IRAM. Usually the baseline was fitted with a second-order polynomial,
but for some spectra we used a third-order polynomial too.

\begin{figure}[!t]
\hspace{-0.1cm}
\includegraphics[width=0.47\textwidth]{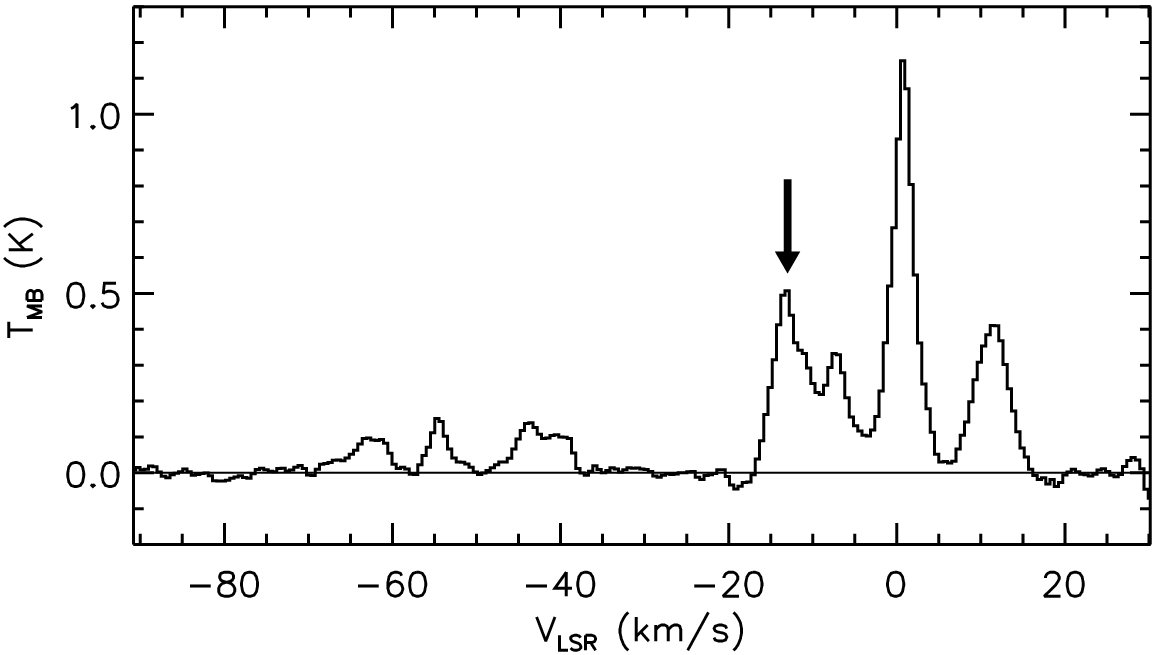}
\caption{SRAO $^{12}$CO $J$ = 1$-$0 average spectrum of 3C434.1. The arrow marks the $-$13 \kms~
component associated with the SNR.
\label{fig2}}
\end{figure}

\section{Results}

\subsection{SRAO $^{12}$CO {\it \textbf{J}} = 1--0}

Several molecular clouds are detected in
the velocity range covered by our observations.
Figure~\ref{fig2} shows the average spectrum of the area:
several velocity components are apparent; bright components
between $-$20 and $+$20~\kms\ and faint components between $-$70~and $-$35~\kms.

\begin{figure}[t]
\includegraphics[width=0.465\textwidth]{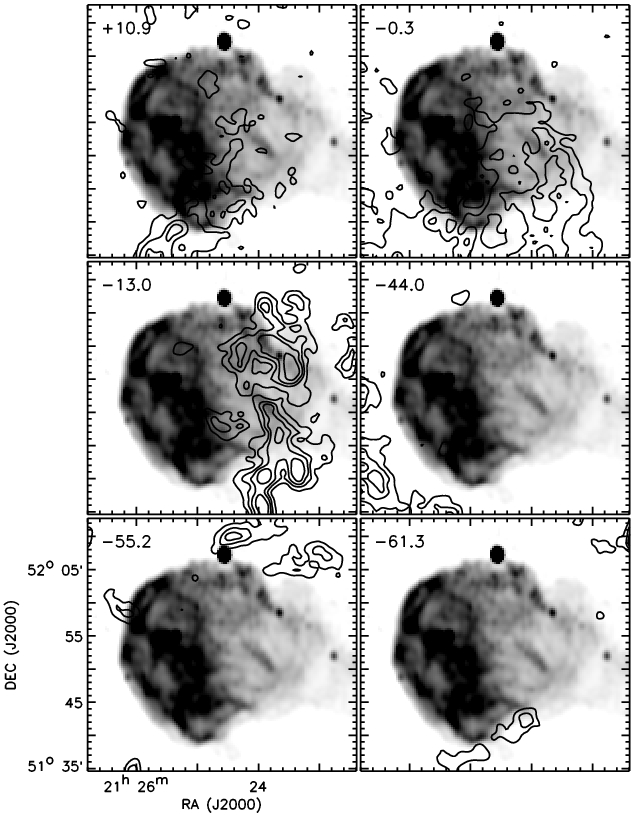}
\caption{SRAO $^{12}$CO $J$ = 1$-$0 contoured channel maps of
prominent velocity components. Individual
frames show single channel maps of width$=0.5$ \kms. Velocity
centers are marked at the top left corners.
The contour levels are 0.9, 2, 3, and
4.5 K in brightness temperature. The gray-scale image in each channel map is the
CGPS 1420 MHz radio continuum image of 3C434.1.
\label{fig3}}
\end{figure}

\begin{figure}[t]
\includegraphics[width=0.465\textwidth]{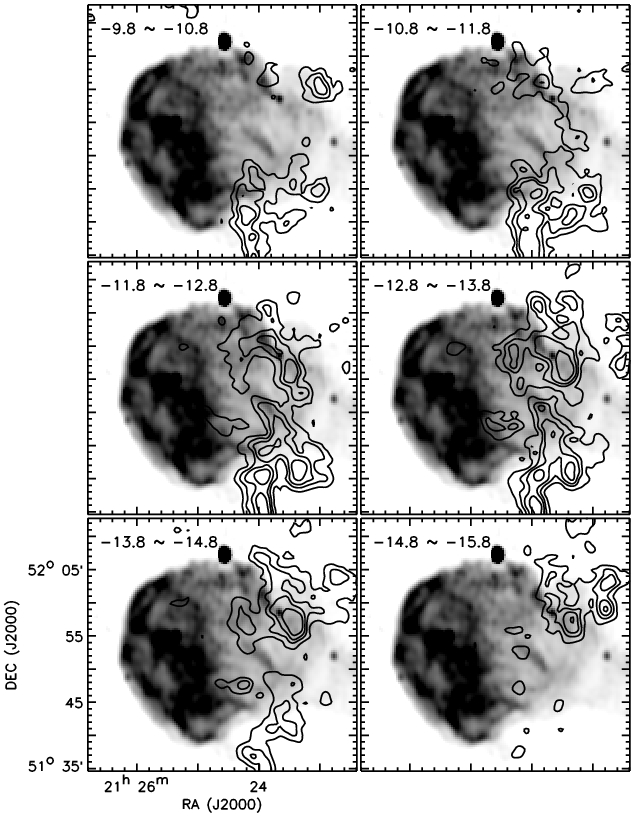}
\caption{Detailed SRAO $^{12}$CO $J$ = 1$-$0 channel maps of 3C434.1 from
$-$10.0 to $-$15.6 \kms.
Each frame shows intensity distributions averaged over two channels or 1.0 \kms.
Contour levels are 0.9, 2, 3 and 4.5 K in brightness temperature.
The Gray-scale is the CGPS 1420 MHz radio continuum image.
\label{fig4}}
\end{figure}

Figure~\ref{fig3} shows the channel maps of the prominent velocity
components.
The components at large negative velocities ($-$70 $\sim$ $-$35~\kms) are
located outside the remnant; the $-$44 \kms~cloud is located a few arc
minutes beyond the southeast boundary. The $-$55 and $-$61 \kms~molecular
clouds are located close to the SNR boundary, but their intensities are very
weak and line widths are
narrow. Thus, we consider it unlikely, that these features have
any relation with
the remnant. The positive velocity (\vlsr~$\geq$ 0~\kms) components are
associated with the complex local clouds of Cep OB2 and Cyg OB7 at 0.8 kpc
\citep{Humphreys78}. These clouds are located in the southern part of the
field, and
we do not find any correlated features with the radio continuum emission.

It is the $-$13~\kms\ cloud that shows a clear spatial correlation with the
radio continuum emission. The cloud is widely distributed over the western
part of the remnant where the radio continuum intensity is low.
Figure~\ref{fig4} compares the distributions of the CO and radio
continuum intensities in detail from \vlsr~= $-$9.8 to $-$15.8~\kms.
The cloud is elongated vertically along the western boundary of the remnant.
At $-$11.8 to $-$12.8 \kms, the southern part of the cloud is bright and there is a thin
filamentary structure that matches well with the western boundary of the torus
structure in the radio continuum. The bright CO region matches well with the southern curved
edge of the radio continuum. At $-$12.8 to $-$13.8 \kms, the northern and southern parts
of the cloud are about equally bright and there is emission along the bar
structure in the radio continuum.

The above morphological correlation between the CO and radio continuum strongly
suggests that the CO cloud is interacting with the SNR and that the
complex
radio morphology of the SNR is caused by this interaction.
We estimate the mass of the western molecular
cloud by integrating the CO emission over \vlsr~= $-$8 to $-$18~\kms\ in an area of
23\arcmin\ $\times$ 39\arcmin\ and assuming a Galactic CO-to-H$_2$ conversion factor
of 2.3 $\times$ $10^{20}$ cm$^{-2}$ (K \kms)$^{-1}$ \citep{Strong88}.
The resulting mass is $(1.00\pm0.02)$$\times10^{4}$$d_{3} ^{2}M_{\sun}$
where $d{_3}$ is the distance to the SNR normalized
to
3 kpc (see \S~4.1)
and the error is $1\sigma$ rms error.

\begin{figure}[!t]
\includegraphics[width=0.465\textwidth]{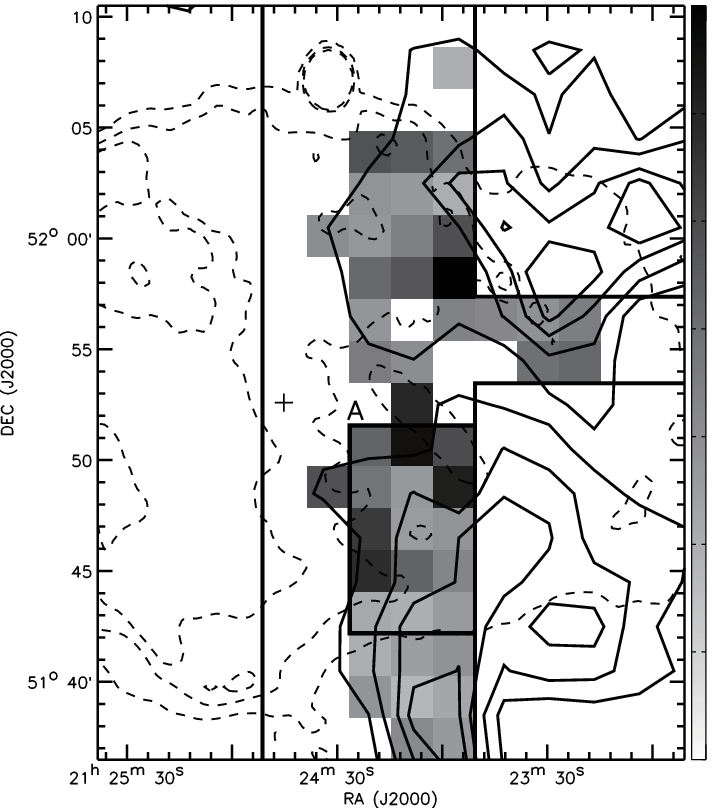}
\caption{$^{12}$CO $J$ = 2$-$1/1$-$0 ratio map in gray scale. The gray scale varies
linearly from 0.1 to 1.8 with the darker scale for the higher ratios.
The dashed contours represent the 1420 MHz radio continuum emission of the SNR (contour levels: 8.2, 10, and
12 K in brightness temperature).
The solid contour is the SRAO $^{12}$CO $J$ = 1$-$0 integrated
intensity map from \vlsr~= $-$16 to $-$8 \kms~(levels: 4, 8, 12, 15 $\times$~1 K \kms).
The bold line marks the area observed with KOSMA $^{12}$CO $J$ = 2$-$1.
\label{fig5}}
\end{figure}

\begin{figure}[!t]
\includegraphics[width=0.465\textwidth]{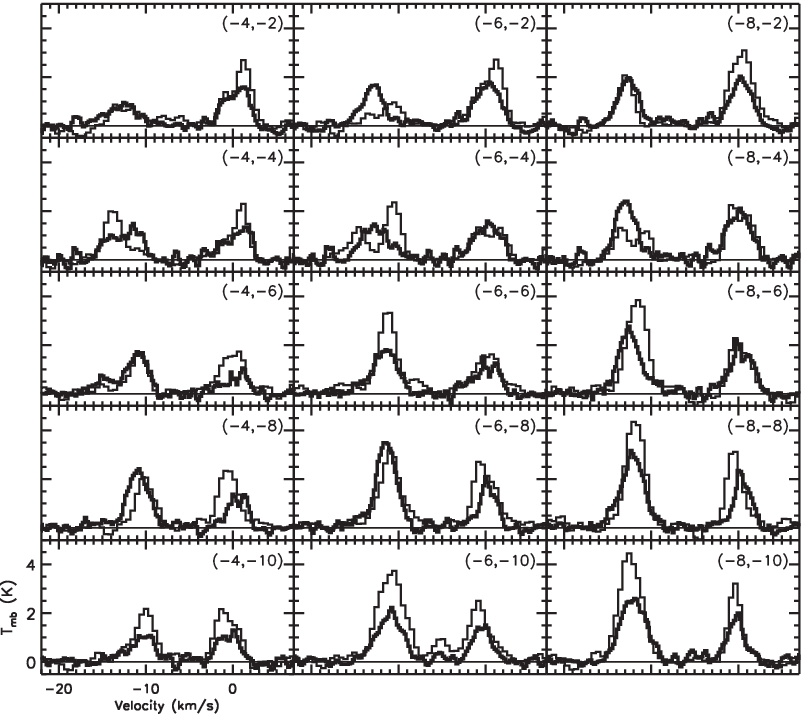}
\caption{CO molecular line spectra of SRAO $J$ = 1$-$0 (thin lines) and
KOSMA $J$ $= 2 - 1$ (thick lines) at region A of Fig.~\ref{fig5}. The positions
of the spectra are marked in each frame as coordinates relative
to the geometrical center of the remnant, (21$^{\rm h}$24$^{\rm m}$45$^{\rm s}$, 51\degr52\arcmin27\arcsec).
\label{fig6}}
\end{figure}

\subsection{KOSMA $^{12}$CO {\it \textbf{J}} = 2--1}

A high
$^{12}$CO ${\it J}$ = 2-1/1-0 ratio
identifies shocked warm and dense CO gas.
Figure~\ref{fig5} shows the spatial variation of the $^{12}$CO ${\it J}$ = 2-1/1-0 ratio obtained
from the SRAO $^{12}$CO $J$ = 1--0 and KOSMA $^{12}$CO $J$ = 2--1 maps integrated over \vlsr~ $= -$16 to $-$8 \kms\ and
binned to 2\arcmin\ $\times$ 2\arcmin. We exclude
pixels with an integrated intensity
less than 2.5 K \kms\ in order to
exclude positions with low S/N ratios from the map.
In Figure~\ref{fig5}, high ($\ge 1.0$) ratios are found toward the southern molecular
cloud. The spectra of this area are shown in Figure~\ref{fig6}.
Note that the 0 \kms\ component shows
${\it J}$ = 2--1 emission weaker than the ${\it J}$ = 1--0 emission with
$^{12}$CO ${\it J}$=2--1/${\it J}$=1--0 ratio of 0.3 to 1.0 which is typical
for giant molecular clouds \citep{sakamoto94}.
The position $(-4, -10)$ also shows an example of ordinary CO line profiles of
unshocked molecular clouds between $-$16 and $-$8 \kms.
Here, the ratio is 0.5 at a velocity of $-$11 \kms.
On the other hand, the $-$13 \kms\ component
has a higher ${\it J}$=2--1/${\it J}$=1--0 ratio. The
$^{12}$CO ${\it J}$=2--1/${\it J}$=1--0 integrated intensity ratio varies from 0.5 to 1.6
in region A in Figure~\ref{fig5} and
the highest value (1.6) is at (21$^{\rm h}$24$^{\rm m}$8$^{\rm s}$,
51\degr50\arcmin30\arcsec) or at $(-6,-2)$ in Figure~\ref{fig6}.
Typically,
the high ratio regions are located at the boundary of
the southern molecular clouds with the peak velocity of $-$13 \kms.
However, at position $(-4, -4)$,
only the $-$11 \kms\ velocity component
shows the high ratio CO line emissions. The ratio is 0.95
at this position using CO integrated intensity results.

\begin{figure*}[!t]
\hspace{0.cm}
\vspace{0.cm}
\includegraphics[width=1\textwidth]{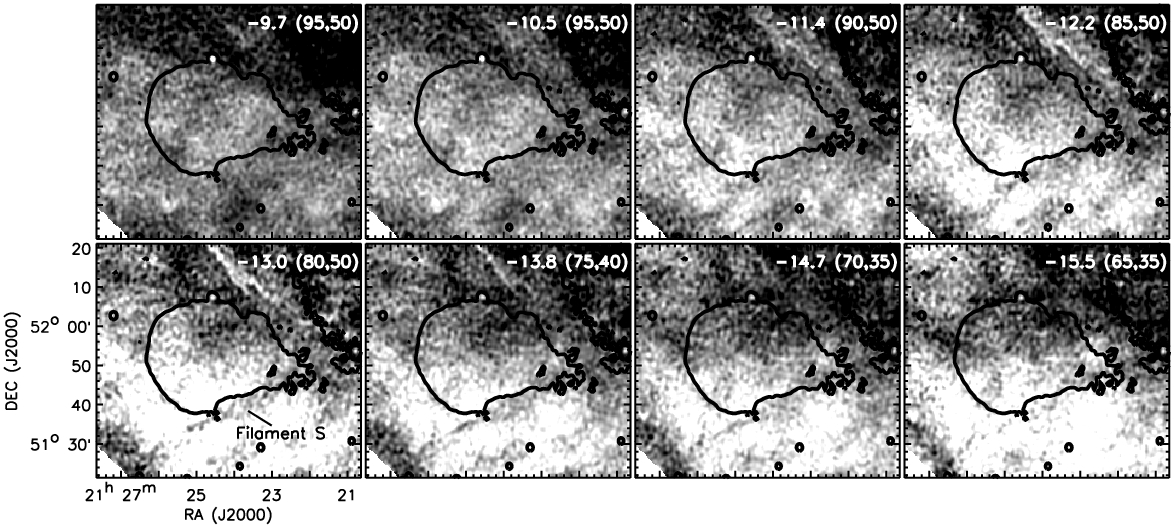}
\caption{CGPS HI distribution of 3C434.1
from $-$9.7 to $-$15.5~\kms\ at 0.8~\kms\ separation.
The gray scales are different for individual frames, and the
maximum and minimum brightness temperatures of the
scales (in Kelvins) are marked in individual frames in parentheses.
The contour is the CGPS 1420 MHz radio continuum contour at 8 K.
\label{fig7}}
\end{figure*}

A ratio higher than $\sim 1$ indicates the presence of warm and dense CO gas, presumably
heated and compressed by the SNR shock \citep[e.g.,][]{Koo01}.
We use the large velocity gradient (LVG) model \citep{Scoville74, Goldreich74} to estimate
the physical parameters of the shocked gas.
Using T$_{mb}$,$_{J=1-0}$~=~0.5 K and $^{12}$R$_{2-1/1-0}$~=~1.6
at the $(-6,-2)$ position,
we obtain \textit{n}(H$_{2}$)~$\sim 2.0 \times 10^{3}$ cm$^{-3}$
and $X(^{12}$CO)/(\textit{dv/dr})~$\sim 7.9 \times 10^{-8}$ pc~(km s$^{-1}$)$^{-1}$ for
T$_{k}$~=~100 K where $X(^{12}$CO)/(\textit{dv/dr}) is
the fractional abundance of $^{12}$CO
relative to H$_{2}$ [$X(^{12}$CO)] per unit velocity gradient interval.
This gives a CO column density of
2.4 $\times$ 10$^{15}$ cm$^{-2}$
assuming a velocity width of $\Delta v$ = 5~\kms.
This is valid when the emitting region
fills the telescope beam (FWHM=$130''$).
If the beam filling factor is lower, e.g., $\sim 0.1$,
the source intrinsic brightnesses and the column density derived scale
up by the inverse factor, i.e. by $\sim 10$,
whereas the density, derived from the line ratio, stays roughly the same.

\section{Discussion}

\subsection{New Distance to 3C434.1 and Its Physical Parameters}

The morphological relation and the high $^{12}$R$_{2-1/1-0}$ ratio strongly
suggest that the SNR 3C434.1 is
interacting with the molecular cloud
that shows CO emission at $-$13~\kms.
The kinematic distance corresponding to $-$13~\kms\ is 3.0 kpc assuming
the Galactic rotation curve of \citet{brand93}
with $R_{0}$ = 8.5 kpc and $v_{0}$ = 220 \kms.
Our kinematic distance is less than
the $4.5\pm 0.9$ kpc distance proposed by \cite{Foster05}.

\cite{Foster04} noted fragmented HI filaments along the SNR boundary at $-$80 \kms\
and interpreted them as parts of a
preexisting stellar wind bubble produced by the
massive progenitor of the SNR.
According to our result, however, the velocity difference is too large
for this component to be associated with the SNR.
We instead examine the correlation of the HI emission at around
$v_{\rm LSR}=-13$~\kms\
with the SNR.
Figure~\ref{fig7} shows the CGPS HI channel maps from $-9.7$ to $-15.5$~\kms.
We first note that there is a large-scale
gradient in HI intensity over the field: The northwestern area is
brighter than the southeastern area. Second, there is a large
filamentary HI-absorption structure in the northwestern area of the field,
but it is not related to the SNR. Absorption from the $-$13 \kms\
cloud interacting with the SNR is not apparent.
Third, there is a long, linear HI structure
just outside of, and along,
the southern SNR boundary at velocities from
 $-$11 to $-$15 \kms\ (labelled "Filament S").
The radio brightness drops sharply across the HI filament.
Therefore, the linearly-shaped southwestern radio boundary
seems to be related to this HI structure.

The geometrical center of the shell is
(21$\mathrm{^h}~$24$\mathrm{^m}~$45$\mathrm{^s}$, 51\degr~52\arcmin~27\arcsec)
and its radius ($R_s$) is 13\arcmin\ or 11.3 pc at 3.0 kpc.
The spherical morphology suggests that we can apply standard
model to this part of the remnant, as has been confirmed
by hydrodynamic simulations \citep[e.g.,][]{tenorio85}.
From the ROSAT X-ray spectra, \cite{Foster05} derived
an electron temperature of $T_e =$ $4.5\times 10^6$~K for
the host X-ray emitting gas, which corresponds to a
shock speed $v_s=(16k_B T_e / 3 \mu)^{1/2}=570$ \kms\
(where
$\mu=1.4m_{\rm H}/2.3=0.61m_{\rm H}$ is the mean mass per particle,
$m_{\rm H}$ = mass of H nuclei, and
$k_B$ is Boltzmann's constant).
This speed is large and we may assume that the SNR is in
its adiabatic phase.
Then, from the Sedov solution,
the radius and the velocity give an age of $0.4R_s/v_s=7,900$~yr.
The estimated volume emission measure is
$\int n_e^2 dV=4.5\times 10^{57}$ cm$^{-6}$ pc$^{3}$ (scaled to 3.0 kpc).
\cite{Foster05} assumed that this X-ray emission comes from a
thin shell in contact with the stellar wind bubble.
We instead assume that the X-ray emitting gas is roughly filling the
eastern half because the X-ray emission is centrally brightened,
which gives an rms electron density of $0.23 $ cm$^{-3}$.
Then the SN explosion energy is, from the Sedov solution,
$E_{SN}=2.1\times 10^{51} n_0 (R/10~{\rm pc})^3 (v_s/10^3~{\rm km~s}^{-1})^2
=0.19 \times 10^{51}$~ ergs, where $n_0(=n_e/1.2)=0.19~{\rm cm}^{-3}$
is the number density of hydrogen nuclei of
the medium taking into account an He abundance of 10\% by number.
This is smaller than the canonical
value of $10^{51}$ ergs but not unreasonable.
The derived parameters of the SNR are summarized in Table~\ref{tb1}.
The dynamical evolution of the SNR including the western part interacting
with the molecular cloud will be discussed in $\S$ 4.3.

\begin{table}[t!]
\begin{center}
\caption{Physical parameters of SNR 3C434.1\label{tb1}}
\begin{tabular}{lc}
\tableline\tableline
Parameter & Value \\
\tableline
Distance (kpc) & 3.0   \\
Radius of eastern shell (pc)     & 11.3  \\
Age (yr)        & 7,900 \\
Ambient hydrogen density (cm$^{-3}$) & 0.19 \\
SN energy (ergs) & 0.19~$\times~10^{51}$\\
H$_{2}$ mass of interacting molecular    & 1.0 $\times$ 10$^{4}$ \\
 cloud ($M_{\sun}$)& \\

\tableline
\end{tabular}
\end{center}
\end{table}

\begin{figure}[!t]
\vspace{0cm}
\includegraphics[width=0.465\textwidth]{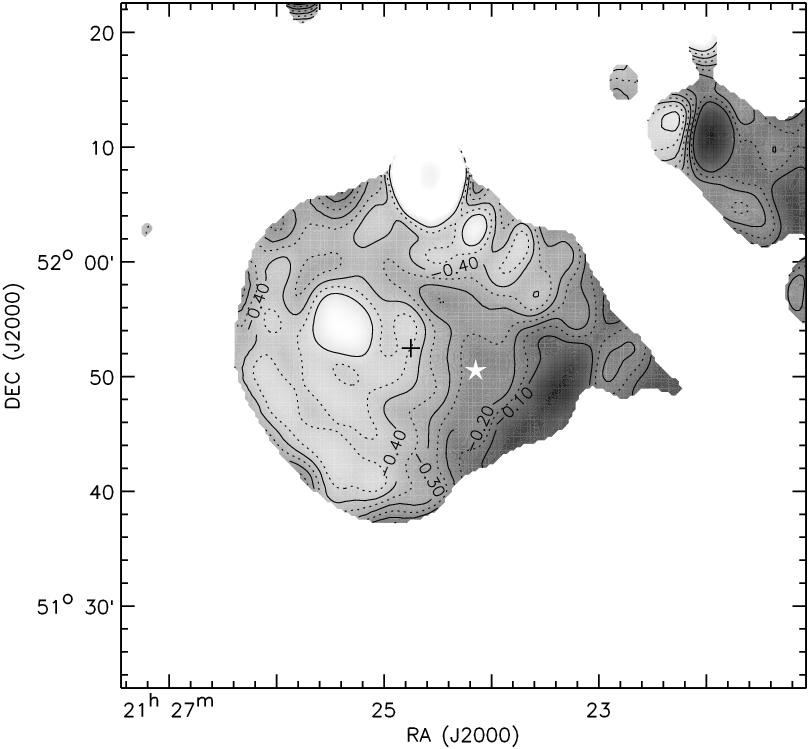}
\caption{Spectral index map of 3C434.1.
The spectral indices ($\alpha$ in $S_\nu\propto \nu^{\alpha}$)
are derived only for the areas with
1420 and 408 MHz brightnesses larger than 1.2 K and 15.2 K corresponding
to 5-$\sigma$ brightnesses, respectively.
The range of gray scale in $\alpha$ is from $-$0.6 to $+$0.2.
The solid contours are drawn
every 0.1 while the dotted contours are drawn every 0.05.
The star symbol shows the location of the high CO ratio region.
\label{fig8}}
\end{figure}

\subsection{Variation of Radio Spectral Index}

SNRs interacting with molecular clouds often show flattening or a turnover of
the radio synchrotron spectrum at low frequencies
around the areas of the interaction, i.e.,
the spectral index becomes larger than the typical value
($\alpha\sim -0.5$; $S_\nu\propto \nu^{\alpha}$)
of shell-type SNRs or even becomes positive.
Some examples are IC 443 \citep{green86}, the Cygnus Loop \citep{Leahy98},
HB 21 \citep{Koo01, Leahy06}, 3C391 \citep{brogan05},
and W44 \citep{castelleti07}.
Several mechanisms have been proposed for the flat radio spectrum:
free-free absorption by thermal plasma in the dense postshock region,
ionization losses from mixed dense molecular material,
second-order Fermi acceleration by the turbulent medium in the postshock
region \citep{Leahy98, ostrowski99}, the higher compression ratio owing to
the energy loss to cosmic ray particles \citep{bykov00}, and the relatively flat
spectrum of ambient cosmic rays entering the shock \citep{chevalier99}.
For 3C434.1, \citet{Foster04} derived a spectral index of $-0.38 \pm 0.03$ from 150 MHz to 2.7 GHz
and noted that the western extension has a slightly flatter spectrum ($\le 0.3$)
than the eastern radio shell. Later \citet{Kothes06} derived a somewhat steeper
index of $-0.48 \pm 0.02$ by fitting the total fluxes from 30 MHz to 5 GHz.

\begin{figure}[!t]
\hspace{-0.3cm}
\includegraphics[width=0.465\textwidth]{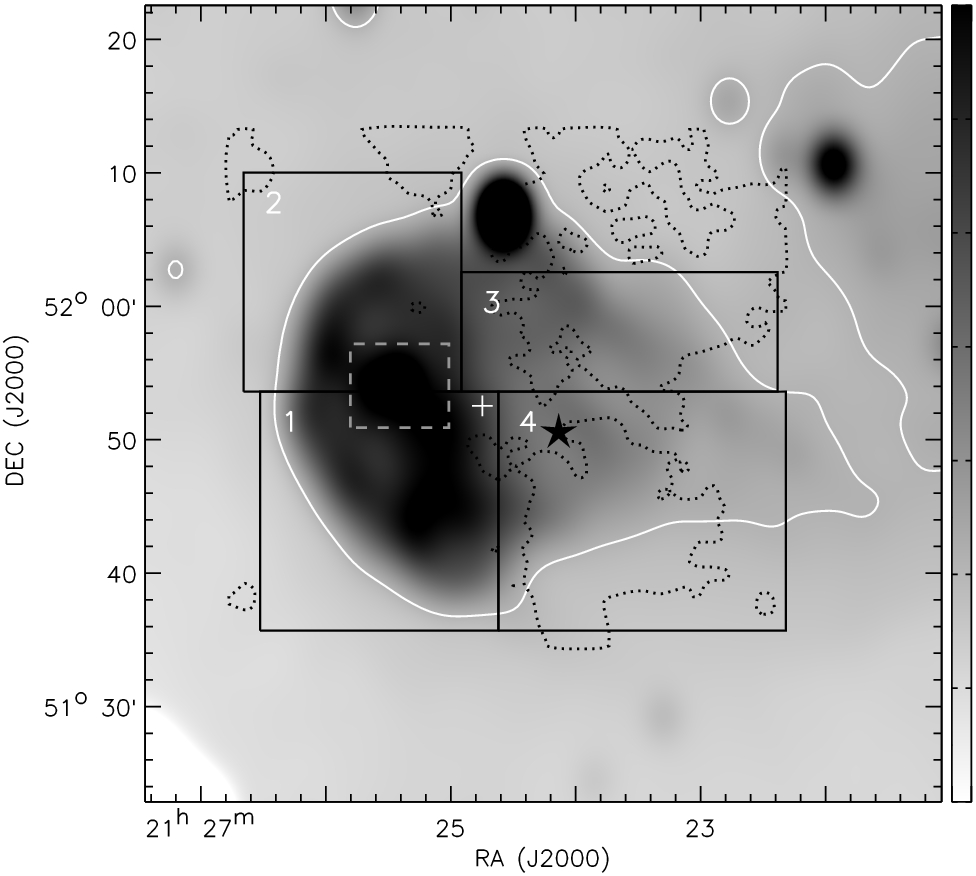}
\caption{Convolved 1420 MHz radio continuum image of 3C434.1.
The Gray scale varies
linearly from 6 to 13 K with the darker scale for the higher temperatures.
Boxes show the regions for
the $T-T$ plots in Fig. 10.
The dashed box represents the area excluded in the $T-T$ plots.
The star symbol shows the location with high
CO $J=2-1/1-0$ ratio.
\label{fig9}}
\end{figure}

We produce a spectral index map using
the CGPS 1420 and 408 MHz radio continuum data.
We first convolve the 1420 MHz image (FWHM=49\arcsec) to the beam size (168\arcsec)
of the 408 MHz image, and we subtract a constant background in each image:
7.3~ and 77 K for 1420 and 408 MHz, respectively.
The resulting spectral index map is shown in Figure~\ref{fig8}, which confirms the flat
spectrum in the western area:
The eastern bright area has an index of $-0.45$ to $-0.3$ whereas the
southwestern area has $-0.1$ to 0.0.
The area around
(21$\mathrm{^h}~$23$\mathrm{^m}~$20$\mathrm{^s}$, 51\degr~49\arcmin~10\arcsec)
has a particular flat spectrum
($\sim 0.0$). The average spectral index of the entire SNR is
$-0.33$.

\begin{figure}[!t]
\hspace{0.cm}
\includegraphics[width=0.465\textwidth]{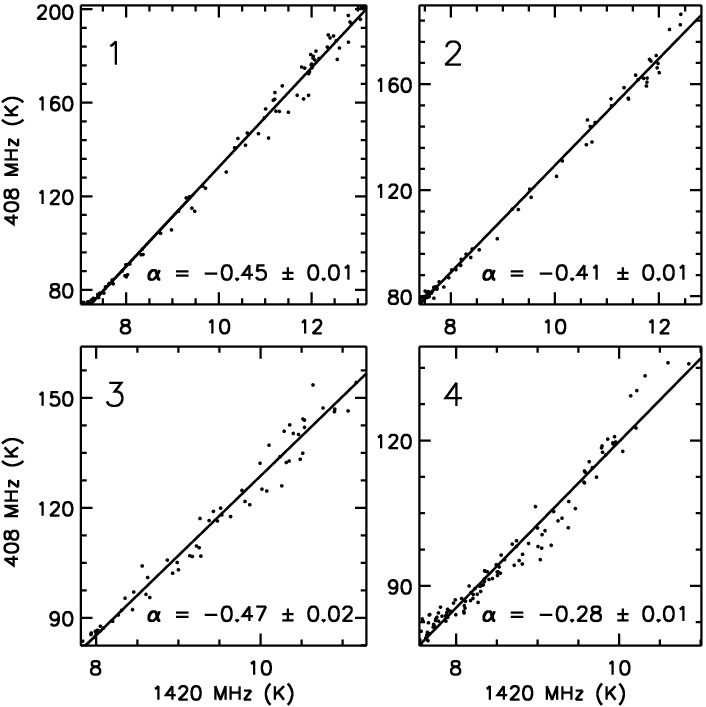}
\caption{$T-T$ plot results of four regions obtained by pixel-to-pixel comparison of
CGPS 408 and 1420 MHz radio continuum images.
\label{fig10}}
\end{figure}

We also derive the spectral index using the ``$T$-$T$ plot method'',
wherein we make a pixel-to-pixel comparison of the
brightness temperatures at two different
frequencies and derive the slope.
The advantage of this method is that the slope
is not affected by potentially
different zero levels of each set of frequency
data and smooth background emission \citep{turtle62}.
We divide the target area
into four regions and derive the index in each region from 1420 and 408 MHz
data.
Figure~\ref{fig9} shows the boundary of the four regions in the convolved
1420 MHz radio continuum image,
and Figure~\ref{fig10} shows their pixel-to-pixel
comparison of the brightness at two frequencies, respectively.
The indices derived from
least-squares fitting are given in the individual graphs.
In three regions, the index varies
little ($-0.47$ to $-0.41$), but in the southwestern area (region 4),
the index is $-0.28\pm 0.01$ which is
considerably flatter than the other regions. This flat spectral index
could be due to the interaction with the molecular
cloud, confirming the reasoning given above.
The average
spectral index of all areas is $-0.42\pm 0.01$.

\subsection{Interaction with the Molecular Cloud and Radio Morphology}

The dynamical evolution of a SNR originating
near the boundary of a dense molecular
cloud has been numerically studied by several authors
\citep[e.g.,][]{tenorio85, wang92, ferreira08}.
The details of the evolution depend on the physical and geometrical parameters
of the simulation, e.g.,
the SN explosion energy, the densities of the molecular cloud and diffuse gas,
the location of SN explosion site, etc.
If the SN explodes in the diffuse medium close to the boundary of the
molecular cloud, the SN blast wave
expands adiabatically until it encounters the cloud.
When the blast wave hits the dense cloud, two waves are produced:
a shock wave propagating into the cloud and a reflected wave
propagating back to the interior of the SNR. The propagating shock wave
decelerates rapidly because of the high density and also because
the reflective wave lowers the driving pressure near the cloud.
Because the molecular cloud's density is considerably larger
than that of the diffuse medium, it is not overly disrupted, and
the SNR appears as a semicircular shell. The semicircular morphology of
the SNR CTB 109, which is another SNR
blocked by a molecular cloud \citep{Tatematsu87, Tatematsu90, Sasaki06},
has been acceptably explained by this scenario \citep{wang92}.

The SNR 3C434.1 also has a semicircular shape, but it does not quite fit into
the above scenario:
It has a narrow, convergent tail that originates
from the western end of the
shell and extends much further ($>25'$ from the center) to the west.
Also, there is a torus that appears to surround the western end of the tail.
The inner boundary of the torus is quite circular with a
clear limb brightening.
This probably indicates that the cloud that is interacting with 3C434.1 is thin and
that the blast wave has broken through it.
This hypothesis is consistent with the presence of an elongated
CO cloud along the western boundary of the SNR.

We have developed a hydrodynamic model to confirm the above hypothesis.
We assume that the SN explodes at 1 pc from the boundary of a
sheetlike cloud
that has a thickness of 5 pc.
The H-nuclei number density of the cloud is assumed to be 20 cm$^{-3}$,
which is a typical density for the interclump medium of giant molecular
clouds \citep[e.g.][and references therein]{chevalier99}.
For the density of the diffuse medium and SN explosion energy,
we adopt 1 cm$^{-3}$ and 10$^{51}$ ergs, respectively.
These canonical values are both about 5 times larger than
those of 3C434.1 but since the radius (and velocity) scales with
$(E_{SN}/n)^{0.2}$ in the Sedov phase,
the resulting morphological evolution in time
should be approximately correct.
The density distribution of a
 core-collapse SN is
characterized by a central region
of approximately constant density and an outer region with a
steep gradient \citep[e.g.,][]{mazner99}.
We assume that the SN ejecta, which are freely expanding,
are composed of a central region
of constant density and an outer region with a steep
power law ($\rho(r)\propto r^{-9}$) with a total mass of 8~\msol.
When the steep power-law portion of the ejecta is interacting
with the ambient medium, a similarity solution
exists \citep{chevalier82}.
Our numerical simulation starts at $t=350$~yr, and the
distributions of the physical parameters at this time
are taken from this similarity solution.
We first considered a model in which
the cloud was shaped like an infinitely extended
disk. This model, however, did
not produce the torus-like structure in the west in time. We therefore
assume that the cloud has a cylindrical hole at the center,
so that the shock can penetrate through the central hole.
The radius of the hole is assumed to be 3 pc.
For the calculation, we adopt the
three-dimensional hydrodynamic code developed by
Harten, Lax and van Leer (HLL) with a
modified cooling effect (\citealt{harten83}).
This HLL code is efficient for describing the shock propagation.
The simulation box consists of $512\times256\times 256$
grid points with a
spatial resolution of 1/16 pc.
We calculate one quadrant column along the symmetric axis ($x$-axis), and
we copy the results to the other quadrants by assuming symmetry.

\begin{figure*}[!t]
\hspace{-0.4cm}
\includegraphics[width=1.01\textwidth]{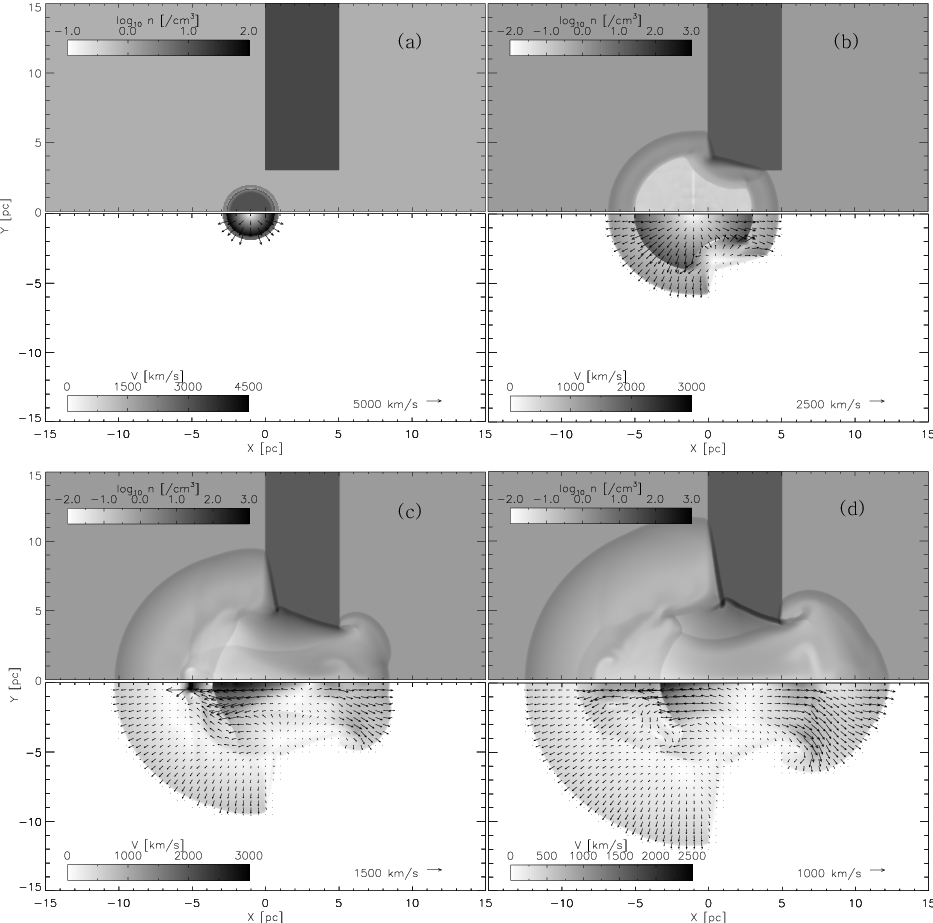}
\caption{Hydrodynamic evolution model for 3C434.1.
Frames from (a) to (d) show the density (upper panels) and
velocity structures (lower) at different stages in the SNR evolution
($t =$ 350, 1950, 4950 and 7950 yr, respectively). In the density panels,
the color bars are shown in logarithmic units. In the velocity panels, arrows
mark both the direction as well as the magnitude of the velocity field.
The molecular cloud with a
cylindrical hole of radius 3 pc is shown in dark gray
on the density panel, but the molecular cloud and intercloud medium cannot be seen
separately in the velocity panel, because their
velocity distribution is assumed
to be static.
\label{fig11}}
\end{figure*}

The results of the numerical simulations are shown in Figure~\ref{fig11}
where we show the time evolution of the density and velocity distributions
in the upper and lower panels, respectively. Figure~\ref{fig11}
(a) shows the initial distributions at $t=350$ yr.
The ambient shock is at 1.89 pc, while the reverse shock
propagating into the SN ejecta is at 1.60 pc from the explosion center.
Between these two shocks, the shocked ambient medium and the shocked ejecta
are separated by a contact discontinuity which is at 1.66 pc \citep[see][]{chevalier82}.
At $t=1,950$~yr (Fig. \ref{fig11}(b)), the SNR blast wave
propagating to the right has been significantly distorted by the
interaction with the cylindrical wall of the cloud.
The cloud boundary has been compressed to high density while
the shock propagating into the dense cloud is considerably decelerated.
Note that the reverse shock in the interacting region
is now propagating backward into the central area in the rest frame, whereas the
reverse shocks in the other regions are
still propagating forward in the rest frame.
At $t=4,950$~yr (Fig. \ref{fig11}(c)), the radius of the left-half of
the SNR becomes 9.4 pc, which is close to that of the Sedov
solution, as expected.
Meanwhile, the blast wave propagating to the right has propagated
through the cylindrical hole and popped out of the cloud
producing a mushroom-like structure.
In the interior, the reverse shocks
from the side walls collide and produce a fast flow to the east.
At $t=7,950$~yr (Fig. \ref{fig11}(d)), which is close to the age of
3C434.1, again the radius of the left-half of
the SNR agrees with that (11.3 pc) of the Sedov solution.
The blast wave on the right by now appears
almost spherical.
Note that the velocity of the blast wave on the right is
larger than
on the left, so that the distance to the rightmost
front from the explosion center is more than 13 pc.
The density and velocity distributions
in the interior of the SNR become complex
owing to the collision of waves and hydrodynamic
instabilities.
Meanwhile, the shock propagating into the
dense cloud became partially radiative, so that the density of the shocked
cloud is very high ($\sim 10^3$~cm$^{-3}$).

\begin{figure}[!t]
\includegraphics[width=0.465\textwidth]{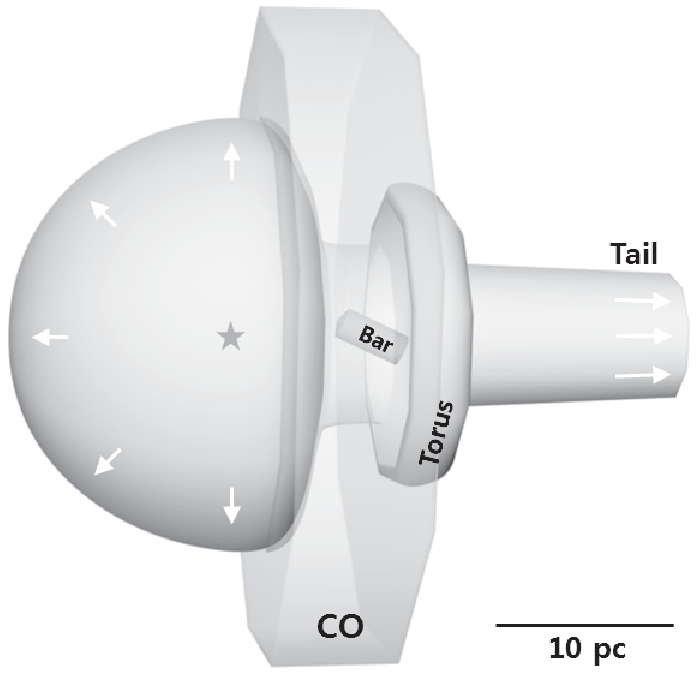}
\caption{Schematic diagram of the SNR 3C434.1. The thin, disklike
molecular cloud interacting with the SNR
is labeled ``CO''. The pronounced radio continuum features
(torus, bar, and tail) are also labeled.
The asterisk symbol marks the position of the SN explosion.
\label{fig12}}
\end{figure}

The resulting morphology of the numerical simulation is close
to the radio morphology of 3C434.1 but there are some differences. In
Figure~\ref{fig12}, we draw a schematic diagram of 3C434.1 including
its radio-continuum features, i.e., torus, bar, and tail.
The spherical eastern half of the SNR with a radius of 11 pc
is quite consistent with the hydrodynamic model.
The torus structure partly matches the mushroom structure
in the model.
The shocked, highly-compressed boundary of the molecular cloud
corresponds to the northwestern SNR boundary with
enhanced radio brightness in the radio map.
The bar structure represents
the dense swept-up cloud
material embedded within the SNR.
The long, radio-continuum tail that
extends much further than the overall radius of the
SNR, however, is not easy to explain.
It seems to indicate that there might have been a
preexisting tunnel of much lower density in the ISM.
An interesting possibility would be
a ``star trail'' as
in the Crab Nebula \citep{blandford83, cox91}.
In the Crab nebula, there is a thin ($\sim 0.5$ pc) jetlike structure in
the radio continuum
to the north, and it was attributed to the trail of the progenitor star;
the progenitor star was moving before the explosion and its
red supergiant wind leaves a trail through which the SN blast wave can
propagate beyond.
In 3C434.1, the radius of the tail is wide,
i.e., 5\arcmin\ or $\sim$~4.4 pc just behind the torus.
However, if the trail is filled with hot, shocked gas, the trail will
expand sideways at about the velocity of the sound speed of the hot gas,
which is 250 \kms\ for $4.5\times 10^6$~K X-ray emitting gas,
the maximum increment of the radius of the tunnel by the SNR would be
$< 2$ pc. Therefore, the tunnel might have had a large radius initially.

\section{Conclusion}

We observed $^{12}$CO ${\it J}$ = 1--0 and ${\it J}$ = 2--1 emission lines toward
the SNR G94.0+1.0 (3C434.1) using the SRAO and KOSMA
telescopes. The presence of a
molecular clouds around the SNR had been known from
the previous low-resolution ($30'$) survey,
but the relation between the two had not been investigated.
Our observations reveal
a thin molecular cloud elongated along the north-south
direction superposed on the western part of the SNR. The spatial correlation of
the CO emission with the radio continuum structure together with
its high $^{12}$CO ${\it J}$ = 2--1/1--0 line ratio strongly suggests that the cloud is
interacting with the SNR. The relatively flat radio spectral index
may also be due to the interaction. The
systemic velocity of the cloud is $-13$ \kms, which yields a
new kinematic distance of 3 kpc to the SNR.
The new distance implies that 3C434.1 is located within the Perseus spiral arm.
This, together with the interaction with the molecular cloud suggests
that 3C434.1 is likely a remnant of
a core-collapse supernova that had a
massive ($\simgt 8$~\msol) stars as
its progenitor.

Our result indicates that the distinct radio morphology of 3C434.1 is
due to the interaction with a dense,
 blocking cloud in the west.
Indeed the spatial correlation between the CO and radio continuum
features is remarkable. Features such as the torus, bar, and tail must
reflect different efficiencies of particle acceleration depending on
the shock speed, ambient density, and magnetic field strength.
Currently,
strong evidence exists for 45 SNRs, which indicates their
interaction with molecular clouds according to the
compilation by \citet{jiang10}, which covers 16\% of the known Galactic SNRs
\citep[see also][]{jeong12}.
3C434.1 is one such
prototypical SNR in this category,
and future high-resolution observations and hydromagnetic
simulations will give us a better
understanding of this intriguing SNR.

\acknowledgments

We would like to
thank Chris McKee and Dave Green for helpful discussions.
Numerical simulations in this paper were performed by using
a high performance computing cluster in
the Korea Astronomy and Space Science Institute.
The KOSMA 3-m submillimeter telescope at the Gornergrat-S\"{u}d
operated
by the University of Cologne in collaboration with Bonn University and
supported by special funding from the Land NRW.
The observatory is administered by the International Foundation Gornergrat and Jungfraujoch.
This research used the facilities of the Canadian Astronomy
Data Centre operated by the National Research Council of Canada
with the support of the Canadian Space Agency.
This work is supported by Basic Science Research program
through the National Research Foundation of Korea (NRF) funded
by the Ministry of Education, Science and Technology (NRF-2011-0007223).

\clearpage



\end{document}